\documentclass[aps,prb,twocolumn,showpacs,groupedaddress]{revtex4}

\usepackage{amssymb}
\usepackage{amsmath}
\usepackage{dcolumn}
\usepackage{bm}
\usepackage{graphicx}
\usepackage{verbatim}

\begin{document}

\title{Unexpected finite-bias visibility dependence in an electronic Mach Zehnder interferometer}

\author{E. Bieri$^1$}
\author{M. Weiss$^1$}
\author{O. G{\"o}ktas$^2$}
\author{M. Hauser$^2$}
\author{S. Csonka$^1$}
\author{S. Oberholzer$^1$}
\author{C. Sch{\"o}nenberger$^1$}
\email{christian.schoenenberger@unibas.ch}
\affiliation{$^1$Physics Department, University of Basel, Klingelbergstr.\,82, CH-4056 Basel, Switzerland}
\affiliation{$^2$Max-Planck-Institut f{\"u}r Festk{\"o}rperforschung, Heisenbergstr. 1, D-70569 Stuttgart, Germany}

\date{\today, submitted}



\begin{abstract}
We use an electronic Mach-Zehnder interferometer to explore the
non-equilibrium coherence of the electron waves within the edge-states
that form in the integral quantum Hall effect.
The visibility of the interference as a function of bias-voltage and
transmission probabilities of the mirrors, which are realized by
quantum point-contacts, reveal an unexpected asymmetry at finite bias
when the transmission probability $T$ of the mirror at the input
of the interferometer is varied between 0 and 100\%, while the transmission probability
of the other mirror at the output is kept fixed.
This can lead to the surprising result of an increasing magnitude
of interference with increasing bias-voltage for certain values of $T$.
A detailed analysis for various transmission probabilities and different
directions of the magnetic field demonstrates that this effect
is not related to the transmission characteristics of a single
quantum point contact, but is an inherent property of the
Mach-Zehnder interferometer with edge-states.
\end{abstract}

\maketitle

\newpage
\section{Introduction}

Interferometers, like the Mach-Zehnder interferometer (MZI),~\cite{Mach-Zehnder,optics_book}
play a decisive role in the foundation of physics. They allow to assess and
quantify the wave nature of light and matter by probing the complex amplitude of
the field.~\cite{Zeilinger1999} The most simple interferometers are so called
two-path interferometers, of which the MZI (Fig.~1) is a particular symmetric
one.~\cite{Mach-Zehnder}
Two path interferometers employ a well collimated incident beam of
light or matter wave generated by source $1$, which is then split by
a partially transmitting mirror $A$ (beam-splitter) with
transmission probability $T_A$ and reflection probability
$R_A=1-T_A$ into two partial beams. After following two different
paths in space, the partial beams are recollected together by a
second half-mirror $B$ forming two output beams that are measured at
detectors $2$ and $3$. Due to particle conservation the two
detectors measure complementary intensities. It therefore suffices
to consider one detector signal. In case of a fully coherent
classical wave with frequency $\omega$, the measured intensity of
the output beam is a periodic function of the difference $\tau$ in
propagation time along the two paths. In the ideal case, the
intensity oscillates between zero and a maximum value, in which case
one refers to a visibility of \mbox{$100$\,\%}.


In recent years, interferometers with a low number of channels have
been implemented in nanoelectronic devices lithographically
fabricated into high-mobility two-dimensional electron gas
systems.~\cite{Hansen2001} In parti\-cuar, MZIs~\cite{Seelig2001}
were realized along these
lines.~\cite{Yacoby1994,Yacoby1995,Yacoby1996,Schuster1997,Buks1998,Ji2003,Neder2006,Litvin2007,Neder2007a,Neder2007b,Roulleau2007,Roulleau2008,Litvin2008}
The two partial beams are either defined by structuring two
pathes~\cite{Yacoby1994,Yacoby1995,Yacoby1996,Schuster1997,Buks1998} or by using the
edge-states,~\cite{Ji2003,Neder2006,Litvin2007,Neder2007a,Neder2007b,Roulleau2007,Roulleau2008,Litvin2008}
that form in a strong magnetic field in the integer quantum Hall
regime.~\cite{Klitzing1980,edge_state1,edge_state2,book_QHE}
In the former approach it is difficult to realize a single quantum channel, because of
residual backscattering at defects induced, for example, by etching.
In contrast, in a strong magnetic field, backscattering is
suppressed leading to the formation of chiral edge-states. The
number of occupied edge-states can easily be controlled through the
magnetic field. In high-mobility samples of moderate densities, one
can even approach a single spin-polarized channel. Edge-states as
electron beams have the additional advantage that ideal tunable
mirrors can be fabricated using quantum point
contacts.~\cite{Henny1999,Oberholzer2006} In these electronic
interferometers the visibility can be measured by sweeping the phase
difference with the aid of the Aharonov-Bohm flux $\Phi$ which can
be changed either by a small variation of the magnetic field or by
a variation of the area inclosed by the two paths employing an electrostatic modulation gate
$V_{mg}$.

Previous experiments of the electronic MZI concentrated
\emph{either} on the energy dependence of the visibility of a symmetric
interferometer ($T_A$\,=\,$T_B$\,=\,0.5) \emph{or} the visibility in
linear response for various transmission probabilities. Here, we
report on the dependence on $T_{A,B}$ \emph{and} energy.

In an interaction free model the total transmission probability
$T_{21} \equiv dI_2/dI_1$ of the MZI consists of two terms:~\cite{Chung2005} an
interference term that depends on the relative phase and a phase
averaged mean transmission probability $\langle T_{21}\rangle$ given by
$T_A T_B + R_A R_B$. In the fully coherent case and without dephasing
the amplitude of the interference term is given by
\begin{equation}
  \hat{T}_{21}=2\sqrt{T_A R_A T_B R_B}~\text{.}
\label{semicircle}
\end{equation}
We define the visibility $\nu$ by the full swing of the
interference signal, i.e. $\nu = 2 \hat{T}_{21}$ (see Fig.~1c).
The visibility is maximal if both mirrors have $50$\,\% transparency,
i.e. $T_A = T_B = 0.5$. If one transmission probability is varied, $\nu$ follows
a semicircle dependence.

Two basic symmetries of the visibility are contained in the above equation Eq.~\ref{semicircle}:
on the one hand, $\nu$ is invariant if $T_A$ is
exchanged with $T_B$ and, on the other hand, if $T_A$ is changed
into $1-T_A$ (or $T_B$ into $1-T_B$). The former more obvious case
states that the outcome of the interference experiment does not
change if input and output are exchanged. This corresponds to time-reversal
symmetry. The latter says that the visibility is symmetric around a
transmission of 50\% for both mirrors separately.
Whereas the first symmetry is respected in our experiment even at finite bias,
we find the second one to be violated. This symmetry breaking,
as we will show below, can lead to the remarkable result of an enhancement of the visibility
with increasing bias-voltage. This is surprising because with
increasing energy one would expect that the number of inelastic scattering channels
increases leading to enhanced dephasing and therefore
reduced visibility. This new feature contrast previous studies
of the energy dependence of the visibility which have always
reported a decay of the interference amplitude with energy (i.e. applied bias voltage).

\begin{figure}[!htb]
\begin{center}
\includegraphics[width=80mm]{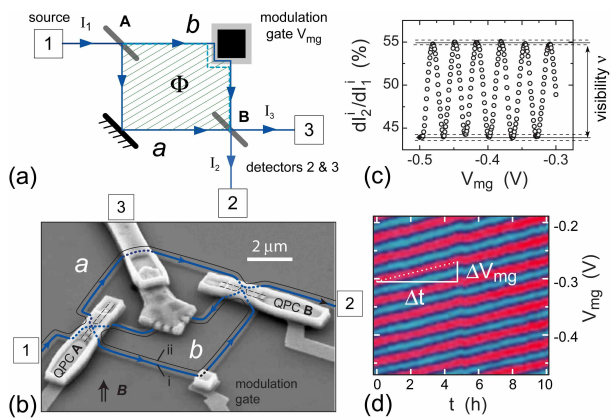} 
\end{center}
\caption{
  (a) Illustration of a Mach-Zehnder interferometer (MZI) with mirrors $A$ and $B$, which
  are characterized by their transmission probabilities $T_A$ and $T_B$. In the electronic version
  a potential $V_{mg}$ at the modulation gate changes the area inclosed by the two partial beams,
  leading to a phase modulation through the Aharonov-Bohm effect.
  (b) Experimental implementation of the MZI in a two-dimensional electron gas.
  The inner ohmic contact and the two metallic split gates which define QPC~A and B are connected
  via free-standing bridges. Of the two edge-states $i$ and $ii$, only the outer one $i$ is partitioned. The
  inner edge-state $ii$ is fully reflected at the two QPCs.
  (c) Due to the consequent constructive and destructive interference, the current intensities $I_{2,3}$
  at the detector contacts $2$ and $3$ oscillate as a function of $V_{mg}$. We measure the differential
  transmission probability $dI^{i}_2/dI^{i}_1$ of the current in the outer edge-state $i$
  and define the visibility $\nu$ as its peak-to-peak modulation as indicated by the
  solid lines in (c). The dashed lines give an indication for the typical measurement error.
  (d) Time-dependence of the phase of the oscillations.
}
\end{figure}

\section{Experimental}

The electronic MZI is implemented in a high-mobility two-dimensional electron gas that forms at the interface of a
GaAs/Ga$_{0.7}$Al$_{0.3}$As-heterojunction \mbox{$120$\,nm} below the surface with an electron
density of 1.6\,$10^{11}$\,cm$^{-2}$ and a mobility of 170\,m$^{2}/$Vs at
\mbox{$4.2$\,K} without illumination. It is defined by wet-chemical etching of a ring structure
combined with two quantum point contacts (QPCs) formed by metallic split gates (Fig.~1b).
Due to the chiral nature of the edge-states, one of the detector contact lies inside the ring.
This current at contact $3$ is drained to ground via a small central ohmic contact.
This ohmic contact as well as the two split gates are connected by
freely suspended bridges, which were realized with a two-layer resist technique employing
PMMA as the bottom and PMMA-MA as the top layer.

The area $A$ enclosed by the two partial beams, which
each have a length of \mbox{$\approx 15$\,$\mu$m}, amounts to $A\approx 38$\,$\mu$m$^2$.
All measurements reported here were carried out in a vertical magnetic field of \mbox{$B =\pm 3.55$\,T}
where two spin-polarized edge-states are present (filling factor two). Of the two
edge-states, denoted as $i$ and $ii$ in Fig.~1b, we only vary the
transmission of the \emph{outer} edge-state, i.e. of channel $i$,
which is the one closest to the sample edge. The inner one is fully
reflected at QPC A and B.
A bias modulation technique is applied to deduce the differential transmission
probability $dI_2/dI_1$ or $dI_3/dI_1$.
A small ac-modulation $V_{ac}$ of \mbox{$1$\,$\mu$V} is superimposed on
the dc-voltage $V_{dc}$ and applied to the source contact $1$.
This contact then injects the current $(e^2/h)V$ into each of
the two edge-states, where $V=V_{dc}+V_{ac}$.
The current at the detector $2$ (or $3$) is measured over
the voltage drop that forms across the detector contact and an additional
ohmic contact connected in series. We then obtain the
differential transmission $dI_2/dI_1$ as the ratio of the
respective ac currents. In this relation
and in all following ones, we implicitly subtract the current
due to the inner edge-state. Hence, $I_{1}-I_{3}$ refer in the following to
the currents in the outer edge-state which we can tune by the QPCs.
A result is shown in Fig.~1c as a function of the voltage $V_{mg}$ on the modulation
gate and in Fig.~1d in addition as a function of time. The gradual
variation of the phase with time at a rate of $\sim$\,$0.33\,\pi/$hour
reflects the decay of the magnetic field of our magnet operated in persistent mode.
This results in a change of the magnetic flux $\Phi$ that threads the interior of the MZI.
Our magnetic field decays with $\approx 18$\,$\mu$T/h, yielding a $2\pi$ phase shift every $6$ hours.
Jumps in the phase are absent which indicates a good quality of the heterostructure.

The peak-to-peak amplitude of the interference, the visibility,
is extracted from the measured data $dI_2/dI_1$, such as the one shown in Fig.~1c, by a
statistical method.~\cite{Roulleau2007} The data points of the oscillating differential current $dI_2/dI_1$,
which was measured as a function of $V_{mg}$ with equidistant increments,
was cast into a histogram. This histogram was then fitted to the form
expected from a sinusoidal dependence for $dI_2/dI_1$ on $V_{mg}$.~\cite{Roulleau2007}
As a control, fast Fourier transformation (FFT) was employed as well.
%
The visibility $\nu$ of the coherent oscillations is in our experiments always smaller than $100$\,$\%$
with a maximal visibility of \mbox{$20$\.\%} at the lowest temperature of \mbox{$35$\,mK}.
As a function of temperature $\theta$ the visibility decays without an indication for
saturation at the lowest temperature, see Fig.~3b. The low temperature decay can be described by an
exponential factor $\exp(-k_B\theta \pi/E_c)$,~\cite{Chung2005} where the characteristic energy
term $E_c$ amounts to \mbox{$2.6$\,$\mu$eV}, corresponding to \mbox{$30$\,mK}.
This value agrees within an order of magnitude with other reports for similar
MZ interferometers. For example, Yang Ji {\it et al.} finds \mbox{$E_c\approx 120$\,mK}~\cite{Ji2003}
and Litvin {\it et al.} have reported \mbox{$E_c = 210$\,mK}.~\cite{Litvin2007}
In the single-particle interference model~\cite{Chung2005} the decay
of the interference with temperature is due to energy averaging over the temperature
window $k_BT$, limiting the temporal coherence of the electron source. In a two-path
interferometer with a path-length difference of $\Delta L$, a wave
with energy $E$ will acquire a phase term in the interference signal given by
$E\Delta L/\hbar v_D$ ($v_D$ is the drift velocity), so that $E_c\approx \hbar v_D/\Delta L$.
Drift velocity for edge-states in the quantum Hall regime have typical values of $10^4$ to
$5\cdot 10^4$\,m/s.~\cite{Komiyama1989} Using the value $E_c$ deduced from the experiment,
this then results in $\Delta L=2.5\dots 12.5$\,$\mu$m. This path-length difference is unreasonably large.
The MZ interferometer has been carefully designed to have the same path length on either
arm. Taking the full arm length of only $15$\,$\mu$m, the estimated $\Delta L$ must be wrong.
The reduced interference amplitude and strong temperature dependence is therefore not caused
by energy self-averaging,~\cite{Roulleau2008} but must be due to another energy-dependent
decoherence mechanism. Such a mechanism may be provided by dephasing due to
electron-electron interaction.~\cite{Seelig2001}

\section{Measurements and Discussion}

\begin{figure}[!htb]
\begin{center}
\includegraphics[width=80mm]{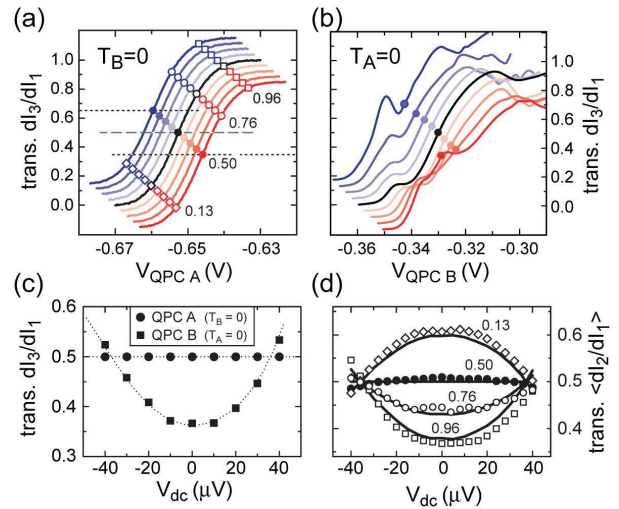} 
\end{center}
\caption{
  (a) Differential transmissions $T_A=dI_3/dI_1$ of the outer edge-state $i$
  through the quantum point-contact QPC~A with QPC~B fully reflective
  as a function of gate voltages $V_{QPC A}$ for different dc bias-voltages
  $V_{dc}$ ranging between $-40$ to \mbox{$40$\,$\mu$V} in steps
  of \mbox{$10$\,$\mu$V}.   The curves are offset for clarity.
  (b), similar to (a), but $T_B$ for QPC~B as the variable and QPC~A fully reflective.
  (c) Energy dependence of the differential transmission $T_A$ of QPC~A at
  \mbox{$V_{QPC A}=-0.65$\,V} and $T_B$ of QPC~B at \mbox{$V_{QPC B}=-0.33$\,V}.
  (d) The dots represent the measured total mean differential
  transmission $\langle dI_2/dI_1 \rangle$ through the MZI as a function of dc bias-voltage
  $V_{dc}$ with the gate voltage of QPC~B fixed to \mbox{$V_{QPC B}=-0.33$\,V}
  and for four different setting of QPC~A corresponding
  to the transmissions $T_A=0.13, 0.5, 0.76$ and $0.96$. The solid lines are calculated
  values using the data in (a) and (b).
}
\end{figure}

Because we intend to vary the transmission probability of both
quantum point-contacts (QPC's) and the dc bias-voltage, the QPC's
have to be carefully characterized. In order to measure the
transmission characteristics of QPC~A (B), the other QPC~B (A) is
completely closed, so that all current is reflected. The current at
contact $3$ is then a measure of the transmission probability.
In Fig.~2a-c we present the differential transmission probability
$T_{A,B}:=dI_{3}/dI_{1}$ of each individual QPC at different
bias-voltages $V_{dc}$ as a function of gate voltage $V_{QPC A,B}$
applied to the QPC's. Whereas $T_A$ is energy independent, $T_B$
reveals a resonance structure at \mbox{$V_{QPC B}=-0.345$\,V}.
In the experiments presented below we varied the transmission probabilities
of both QPC's over a large range. Once both point contacts are
adjusted the transmission probability of each individual QPC cannot be measured anymore.
We then only have access to the mean total transmission probability $\langle T_{21} \rangle $
of the MZI between contact $1$ and $2$ (or $1$ and $3$).
In Fig.~2d we compare the measured mean transmission probability $\langle dI_2/dI_1\rangle$
(symbols) with the expected value $\langle T_{21} \rangle = T_A T_B + R_A R_B$ (solid curves)
using the transmission and reflection coefficients $T_{A,B}$, $R_{A,B}$,
experimentally determined before. The good agreement show that we are able
to adjust the two QPCs independently in the closed interferometer and even for $V_{dc}\not = 0$.

\begin{figure}[!htb]
\begin{center}
\includegraphics[width=80mm]{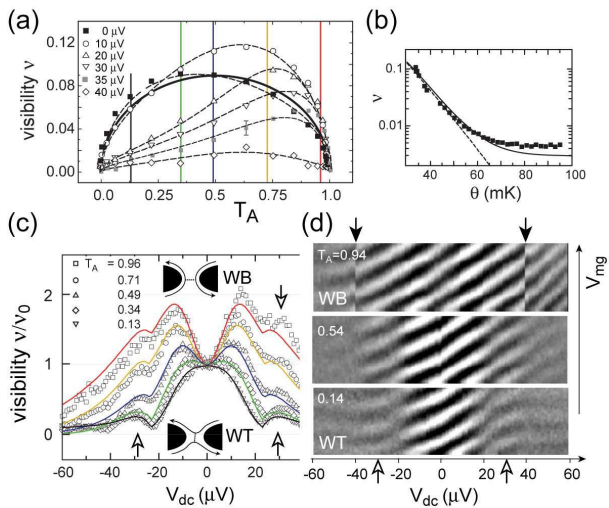} 
\end{center}
\label{fig3}
\caption{
  (a) Visibility $\nu$ for six different dc bias-voltages as a function of
  the transmission probability of QPC~A. The gate voltage on QPC~B is fixed
  ($T_B=0.56$ at $V_{dc}=0$).
  (b) As a function of temperature $\theta$, the visibility follows
  an exponential decay which when fitted to the dependence
  $\propto exp(-k_B\theta\pi/E_c)$ yields for the characteristic energy scale
  $E_c$ a value of $2.6$\,$\mu$eV, corresponding to $30$\,mK.
  (c) The dependence of the visibility on the dc bias-voltage $V_{dc}$
  scaled to the zero-bias value $\nu_0$ for five different $T_A$ values
  ($T_A=0.96$, $0.71$, $0.49$, $0.34$, and $0.13$) and QPC~B fixed as in (a).
  In the weak backscattering (WB) limit the visibility first grows with increasing $V_{dc}$, whereas
  it decays in the opposite case of weak tunneling (WT).
  The curves in (a) and (b) are guide to the eyes.
  (d) The phase evolution is visible in the measured differential transmission
  through the MZI as function of $V_{dc}$ in the WB, WT and an intermediate regime.
}
\end{figure}

In Fig.~3a the peak-to-peak visibility $\nu$ deduced from the differential transmission
probability in the outer edge-state $dI_2/dI_1$ is shown as a function of the transmission probability
of the first beam-splitter at the entrance of the MZI, defined by
QPC~A with QPC~B fixed ($T_B = 0.56$) for different dc bias-voltages
$V_{dc}$. According to theory, $\nu$ should be proportional to
$\sqrt{T_AR_A}=\sqrt{T_A(1-T_A)}$. This dependence is shown as a
solid curve. Taking the error bar of the experiment into account, a
good agreement is found for the zero bias case. In the
non-equilibrium case, for bias-voltages \mbox{$V_{dc}\gtrsim
10$\,$\mu$V} a striking asymmetry appears. This asymmetry is
in\-consistent with the relation $\nu \propto \sqrt{T_AR_A}$, which is
symmetric around $50$\,\% transmission. As compared to the
equilibrium values and for not too large bias-voltages, the
visibility increases with $V_{dc}$ for large $T_A$ values, whereas
it always decays for small ones. This is better visible in Fig.~3c
which shows the scaled visibility $\nu/\nu_0$ (where $\nu_0$ is the visibility
at $V_{dc}=0$ ) for five different settings of $T_A$ running
from top to bottom from the weak-backscattering (WB)
limit in which $T_A$ is large to the weak tunneling (WT) limit in which $T_A$ is small.
In contrast to the WB limit, $\nu/\nu_0$ first decays with increasing $|V_{Ddc}|$ to develop
a side lobe appearing symmetrically approximately at $\pm 30$\,$\mu$V
on either bias side. The side lobes are visible in all curves in Fig.~3c (open arrows),
as well as in the grey-scale phase image Fig.~3d.
Such a dependence of $\nu$ on $V_{dc}$, a central lobe accompanied
by side lobes, have been reported in several
publications.~\cite{Neder2006,Litvin2007,Neder2007b,Roulleau2007}
In some reports only one pair of side lobe appears, whereas in other studies
several were observed. Though only one pair of side lobe is clearly visible
in Fig.~3c, the grey-scale image in Fig.~3d suggests the appearance of a second pair
at \mbox{$\approx 50$\,$\mu$V}.
In addition, it has been reported that the
phase of the interference pattern in a MZI can jump by $\pi$ on the
transition from one lobe to the other.~\cite{Neder2006,Neder2007b}
Bias-induced $\pi$-phase jumps where reported before in
Aharanov-Bohm rings with tunneling barriers~\cite{vOudenaarden1998,vdWiel2003}
and interpreted as the electric Aharanov-Bohm effect.
Whereas the phase is seen in Fig.~3d to gradually evolve
at small bias-voltage, a more complex curved pattern is present at larger
bias-voltages.~\cite{vdWiel2003,Leturcq2006} However, in the WT
limit, the lobe structure does not seem to be accompanied by a phase
jump of $\pi$ as reported by Neder {\it et al.}~\cite{Neder2006,Neder2007b}
In contrast, clear phase jumps appear in the WB limit as seen in Fig.~3d (filled arrow).
Generally, the visibility dependence in the WB limit
strongly contrast the WT one. In the former $\nu$ is
\emph{enhanced} for small bias-voltages above the equilibrium value.
At larger bias-voltages $\nu$ decays displaying a weak modulation
best visible in the grey-plot of Fig.~3d. The reversed dependence
on $V_{dc}$ for small bias-voltages in the two limits shows,
that these limits are inequivalent in the MZI with
edge-states. This is an effect, which one would not anticipate in a
model in which the two partial beams are equivalent and isolated
from the environment. Before discussing the origin of this effect,
we have to prove that the effect is not induced by \emph{one} of
the QPC, but is a property of the interferometer.

The 
discovered asymmetry in the visibility is observed in Fig.~3a as a
function of the transmission probability $T_A$ of QPC~A, which is
the first beam-splitter at the entrance of the interferometer. One could
argue that this effect may be a property of this particular quantum
point-contact. At first, this seems unlikely because QPC~A is the
point-contact that shows a very smooth and energy-independent
transition from $T_A=0$ to $T_A=1$ (Fig.~2a). Secondly, we
can revert the role of the two QPC's by changing the direction through
which the current flows. This is achieved by switching the magnetic
field and placing the source at contact $2$. We then measure the differential
transmission $T_{12} \equiv dI_1/dI_2$ from contact $2$ to the drain
contact $1$, instead of $T_{21}$. Now, QPC~B is the first beam-splitter in the
interferometer and QPC~A the second one. Hence, the asymmetry in $\nu$
should show up if we now vary QPC~B. This is indeed the case
as demonstrated in Fig.~4a. If on the other hand we vary the
transmission probability of the second QPC, QPC~A, a symmetric
dependence of $\nu$ is found. This is shown in Fig.~4b.

\begin{figure}[!htb]
\begin{center}
\includegraphics[width=85mm]{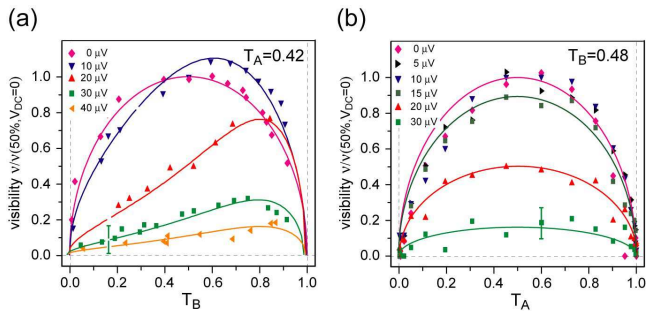} 
\end{center}
\label{fig4}
\caption{
  (a) Visibility $\nu$ normalized  to the value observed at $T_{A,B}=0.5$ and
  $V_{dc}=0$ for different dc bias-voltages (different labels) as a function of
  the transmission probability of QPC~B with QPC~A fixed and vice versa in (b).
  In these experiments the sign of the magnetic field was reversed, as well as the role
  of source and drain contacts. The dc-bias voltage is now applied to contact $2$, while contact
  $1$ is used to measure the current. The asymmetry observed in Fig.~3a and the present figure (b)
  is determined by the first QPC, but \emph{not} by the second one.
}
\end{figure}

Our results clearly show, that the beam-splitter at the entrance determines the
asymmetry and that this is independent on the global direction taken
by the current through the interferometer. Contrary to this
asymmetry, the dependence on the exit beam-splitter reveals the
well-know symmetric semicircle dependence. This proves that the
observed asymmetric visibility dependence is a property of the
interferometer, and not of one of the individual QPC's.

In order to understand this effect, we have to look for a property
that is different for the two partial beams. The first beam-splitter
divides the incident beam into two portions that propagate further
along the two different paths. In the schematics of Fig.~1a the two
paths are alike (except for the modulation gate). In the real device
shown in Fig.~1b, the two trajectories bear an additional difference
which is evident if one recalls that we are working at filling factor $2$.
The second inner edge-state \textit{ii}, which we have ignored in the previous discussion,
is differently occupied in arm \textit{a} and \textit{b}, see the illustration
in Fig.~5 (see also Fig.~3 of Ref.~[31]). In arm \textit{a} it is filled up to
the potential $V$ of the source contact, i.e. it carries the full
quantum current $(e^2/h)V$. In contrast, in arm \textit{b}
it is kept at zero potential, because contact $3$ is
connected to ground. In terms of state occupation there is an
asymmetry between the two arms. This asymmetry is smallest
in the WB limit, where on each arm one edge-state
is filled up to $V$ while the other remains `empty' (filled to zero
potential). The asymmetry is largest in the opposite case of WT.
In arm \textit{b} both edge-states are then filled closely
to zero potential, whereas they are both filled to $V$ in \textit{a}.
It is in this regime in which the visibility is strongly
suppressed, hence, in which dephasing is large.
The experiment therefore shows that the pair of
edge-states are most strongly susceptible to dephasing if their Fermi edges are
close together. Because dephasing can efficiently be mediated by electron-electron interaction
in a low-dimensional system, one could reason that the two edge-states have
similar group velocities at similar filling, leading
to similar charge-density wave excitations. The coupling of such plasmonic
charge excitation may strongly influence the coherence in the
interfometer.~\cite{Sukhorukov2007,Levkivskyi2008}

\begin{figure}[!htb]
\begin{center}
\includegraphics[width=80mm]{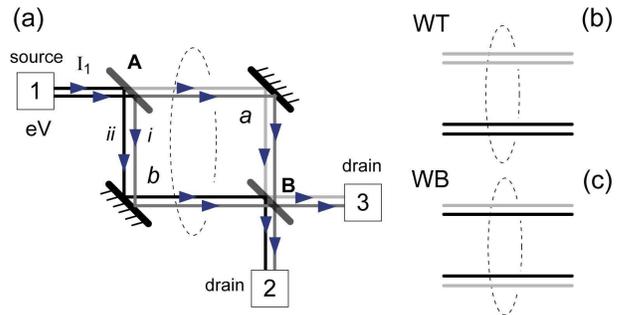} 
\end{center}
\label{fig5}
\caption{
  (a) Illustration of the MZI showing the two transport channels.
  The dark black lines indicate edge-states with filling up to $eV$, whereas
  the grey one have a intermediate filling. Light grey indicate low filling.
  (b) and (c) display the fillings in the two interferometer arms in
  the weak-tunneling regime (WT) and the weak back\-scattering regime (WB), respectively.
  See also Fig.~3 of Ref.~[31].
}
\end{figure}

That the edge-state are strongly coupled electro-statically is reflected
in the phase evolution. The phase in Fig.~3d gradually shifts with bias voltage $V_{dc}$.
This shifts amounts to $2\pi$ for approximately \mbox{$20$\,$\mu$V}. We can understand this as
an electro\-static gating from the inner edge-state onto the outer one. The inner edge-state
\textit{ii} in the outer arm \textit{a} of the MZI is also biased to $V_{dc}$ and electro\-statically influences
the chemical potential in the outer edge-state \textit{i}, which is the one taking directly part in
the interference. The electro\-static phase is then given by $\alpha 2 e V_{gate}L/{\hbar v_D}$,
where $V_{gate}$ equals $V_{dc}$, $v_D=10^4-5\cdot 10^4$\,m/s is the drift velocity,~\cite{Komiyama1989}
$L=15$\,$\mu$m the arm length of the interferometer, and $\alpha$ the gate-coupling efficiency.
With the measured values, our experiment is consistent with $\alpha = 0.14 \dots 0.75$ depending
on the exact (but unknown) drift velocity. Because of the close-proximity of the edge-states,
a large coupling is plausible. The strong electric coupling between the edge-states may provide
a channel for dephasing as proposed by Levkivskyi and Sukhorukov.~\cite{Levkivskyi2008}
In their theory the excitations are dipolar and charged edge magneto\-plasmons.
This theory results in a dephasing rate which is inversely
proportional to the temperature $\theta$, which was confirmed recently.~\cite{Roulleau2008}
A similar dependence was also derived for a single channel when screening was taken into
account in a self-consistent manner.~\cite{Seelig2001} In the latter model, dephasing is caused
by intrinsic phase fluctuations, driven by the thermal bath. A related concept
of intrinsic dephasing in a single channel has been put forward in two other papers
recently.~\cite{Neder2008,Young2008}

While these theories provide a mechanism for the decay of the measured visibility with
$\theta$ and bias voltage $V_{dc}$, we also observe an increase in $\nu$ as a function
of $V_{dc}$ for certain settings of the QPCs. This can qualitatively be understood using a
simple argument. We assume in the following vanishing dephasing. The detector current in the
outer edge-state $I_2$ as a function of the bias voltage $V$ can then be written as
\begin{equation}
  I_2(V)=\frac{e^2}{h}V\left(\langle T_{21}(V)\rangle + \hat{T}_{21}(V)cos(\phi_0 + \beta V_{mg} + \gamma V) \right)
\end{equation}
where $\langle T_{21}(V)\rangle$ is the mean transmission, $T_{21}(V)$ the interference amplitude,
$\phi_0$ a static phase term, depending on the Aharonov-Bohm flux, and $\beta$ and $\gamma$ parameters
that describe the coupling from the modulation gate and the inner edge-state onto the phase.
With the previous notation $\gamma = \alpha 2L/\hbar v_D = 0.3$\,($\mu$V)$^{-1}$. Because we have measured
the differential transmission we have to evaluate $dI_2/dI_1=(h/e^2)(\partial I_2/\partial V)$. It contains
two interference term: $(\hat{T}_{21}+V\partial \hat{T}_{21}/\partial V)cos(\dots)$ and
$-V\gamma \hat{T}_{21}V\gamma sin(\dots)$, resulting in a visibility of
\begin{equation}
  \nu = 2 \sqrt{(\hat{T}_{21}+V\partial\hat{T}_{21}/\partial V)^2+(\hat{T}_{21}V\gamma )^2}.
  \label{equ_vis}
\end{equation}
Even for ideal beam-splitters, for which $\hat{T}_{21}$ does not depend on $V$, the visibility
can increase with bias voltage $V$ due to the term $\hat{T}_{21}V \gamma$.
If we assume $\partial\hat{T}_{21}/\partial V=0$ and compare the zero-bias
curvature of the $\nu/\nu_0 (V_{dc})$ in Fig.~3c (WB curve) with Eq.~\ref{equ_vis}, we deduce for
$\gamma$ a value of $0.15$\,($\mu$m)$^{-1}$ which is in reasonable agreement with the value
deduced from the phase shift before, i.e. $\gamma = 0.3$\,($\mu$V)$^{-1}$.
If we add in addition an exponential dephasing term with a dephasing rate
proportional to $V$,~\cite{Seelig2001,Levkivskyi2008}
i.e. $\hat{T}_{21} = T_0 exp(-V/V_{\phi})$, the visibility at small voltages $V << V_{\phi}$
will approximately follow $1-V/V_{\phi}+(V\gamma)^2/2$ and be dominated by the
exponential decay for $V \agt V_{\phi}$. This shows that $\nu$ can indeed grow for not too large
applied voltages, provided that $V_{\phi}$ is large, i.e. $V_{\phi}> 2/V\gamma^2$.
What theory has to provide is the relation between channel occupation determined by
the mirror settings and the parameters $\gamma$ and $V_{\phi}$ to understand
the peculiar asymmetry in $\nu(T_A)$ of Fig.~3a and $\nu(T_B)$ of Fig.~4a.

\section{Conclusion}

Mach-Zehnder interferometers were proposed as building blocks for the realization
of orbitally entangled states.~\cite{Samuelson2004,Neder_HBT_2007}
In order to prove the entanglement, however, one would need to
analyze a Bell inequality, for which the probability for
two-particle coincidences are expressed in terms of zero-frequency
current-noise correlators. Such measurements are performed out of
equilibrium and for many different transmission probabilities of the
quantum point contacts. An understanding of the visibility at finite
bias~\cite{Neder2008,Young2008} and for different transmission
probabilities is therefore crucial for future experiments along this line.
On its own, the observed intrinsic asymmetry in the visibility of
a MZI with edge-states, yields new insights
in the properties edge-states, in particular in their phase coherence
and mutual coupling. It would be interesting to see whether the recent
theories~\cite{Levkivskyi2008,Neder2008,Young2008}
are able to reproduce the asymmetric dependence of the visibility on
the mirror setting at the input of the MZI.

\section{Acknowledgement}

{\noindent\small{\bf Acknowledgments}
  The authors wish to thank  M.\ B{\"u}ttiker, W. Dietsche, I.\ P.\ Levkivskyi, C.\ Strunk, E.\ V.\ Sukhorukov
   P. Roche, J. Weis and D. Zumb{\"u}hl. This work was supported by the Swiss National Science
  Foundation and the NCCR on Nanoscale Science.


\end{document}